# Quenched Staggered Spectrum at $\beta$ =6.0, 6.2 and 6.4 *

Gregory Kilcup[a]

[a]Department of Physics, The Ohio State University, 174 West 18th Ave, Columbus, Ohio 43210

We give a preliminary report on the hadron spectrum on an ensemble of quenched lattices at $\beta$ values of 6.0, 6.2 and 6.4, using staggered fermions and collaborators Rajan Gupta and Steve Sharpe. Because of the relatively small number of configurations we pay marginally more attention to the question of statistics than in previous analyses. We are unable to discredit quenched QCD.

## 1. The Ensemble

As part of our ongoing program in weak matrix elements, we have accumulated a set of quenched gauge configurations and staggered quark propagators at several different $\beta$ values. As a first step in our analysis we have computed the hadron spectrum. For the results on the scaling of matrix elements, see elsewhere in these proceedings.

The parameters of the simulations are listed in Table 1. The configurations were generated principally using the overrelaxed algorithm, with one 20-hit Metropolis step out of every 10 sweeps. Staggered quark propagators were computed using periodic boundaries in space, and both periodic and antiperiodic in time. Taking linear combinations, we formed forward- and backward-going propagators. These propagators are exactly the same as would result from doubling the lattices in the time dimension, but this method saves memory space. For each quark mass and boundary condition we computed four propagators, using two types of wall sources (with phases 1 and $(-)^n$) and two source timeslices (0 and $N_t/2$). The wall source was defined by fixing to Landau gauge, as this was convenient for matrix element calculations.

## 2. Curve Fitting

Having measured the average value of some correlator $\overline{G}(t)$, and chosen some fit function $f(\alpha; t)$, the generic curve fitting problems are (1) to determine parameters $\alpha$ which yield the best fit, and (2) to decide if the fit is reasonable or not.

Table 1
The Statistical Ensemble

| Beta: | 6.0 | 6.2 | 6.4 |
|---|---|---|---|
| Volume: | $16^3 \times 24$ | $32^3 \times 48$ | $32^3 \times 48$ |
| $N_{\text{samp}}$: | 41 | 23 | 24 |
| $m_{\text{quark}}$: | .03 | .025 | .015 |
| | .02 | .015 | .010 |
| | .01 | .010 | .005 |
| | | .005 | |

The standard method in use in the lattice community[1] is to solve both problems at once by estimating the covariance matrix from the data itself, and then minimizing the usual goodness-of-fit statistic

$$\xi \equiv \delta G \cdot C^{-1} \cdot \delta G. \tag{1}$$

Here $\delta G$ is the vector of deviations between the data and the model curve, while $C(t, t')$ is the estimated covariance matrix. In the limit of large sample size, the fluctations in the average value of a correlator converge to a Gaussian distribution described by the covariance matrix, and the statistic $\xi$ is distributed as a $\chi^2$. One can then test the hypothesis that $f(\alpha; t)$ is the true infinite sample mean of value of $G(t)$ by asking how often a random Gaussian fluctuation away from this hypothetical true mean would lead to a value of $\xi$ as large or larger than the one observed. This fraction is the confidence level of the fit, which is readily computable with standard routines. Typically one demands a confidence level of at least .05 to trust a fit at all.

For finite sample size $N$, one may proceed by assuming that the fluctuations $\delta G$ about the true

---

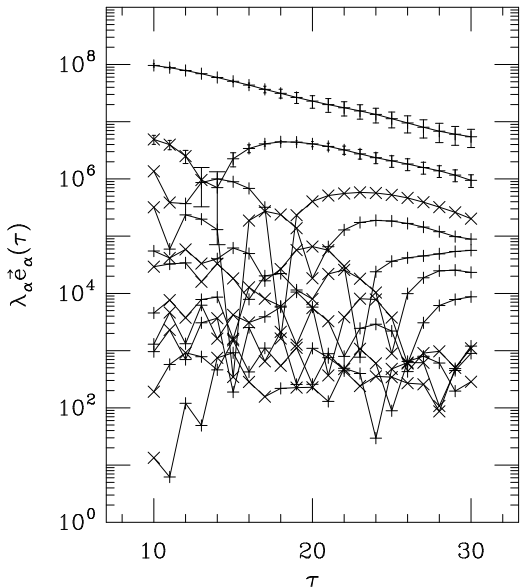

Figure 1. Modes of the covariance matrix for the $\pi$ channel at $\beta = 6.4$ and $m_q = .015$. Plusses and crosses indicate positive and negative values.

$N$ eigenmodes of the covariance matrix. These eigenmodes are notoriously ill-determined. In figure 1 we show the first 11 modes of the covariance matrix for one of our pion correlators, scaling each unit-normalized eigenvector by the corresponding eigenvalue. As expected, the dominant mode of fluctuation is the one where all timeslices oscillate up and down together. The error bars are determined by jackknife, and are drawn on only the first two modes, since beyond that they become huge. The most ill-determined modes have eigenvalues several orders of magnitude smaller than the first mode, and these then dominate the fit. In figure 2 we resolve a typical uncorrelated fit mean are at least approximately Gaussian, and that $\xi$ is approximately a $\chi^2$ statistic. More precisely, if we fit to $T$ timelices and have a sample of $N$ gauge configurations with $N > T$, then $\frac{N-T}{N-1}\xi$ is distributed as an $F$-statistic with $T$ and $N-T$ degrees of freedom[2]. (If $x_1$ and $x_2$ are distributed as $\chi^2$ with $n_1$ and $n_2$ degrees of freedom respectively, then by definition $y \equiv \frac{x_1}{x_2}$ follows an F-distribution with $n_1$ and $n_2$ degrees of freedom.) For large $N$ this of course reduces to an ordinary $\chi^2$, but for $N$ of order $T$ (which it always is for us), the F-distribution has a long tail. For samples small enough that $T > N$ (as is too often the case), the covariance matrix is singular, and one typically proceeds by defining $\xi$ in terms of the singular value decomposition (SVD) inverse. In either case, treating $\xi$ as an ordinary $\chi^2$ leads to "peculiar" fits where the fit curve misses all the data points, and gives misleading confidence levels.

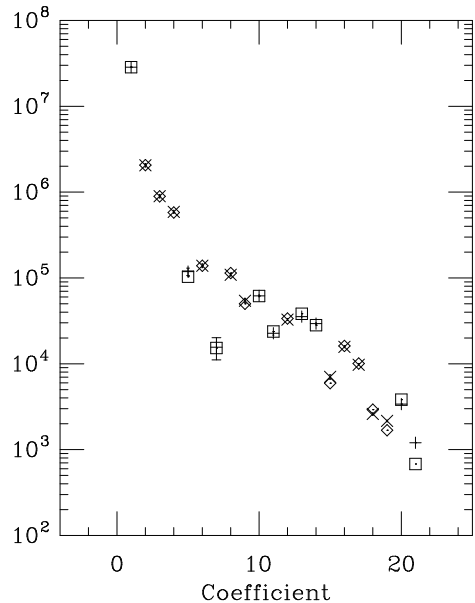

Figure 2. Resolution of a two particle fit in the space of (unit-normalized) eigenmodes. Plusses and crosses indicate positive and negative values of coefficients for the data; squares and diamonds show the fit function.

The root of the problem is that with $N$ samples, one cannot really hope to reliably determine in the space of eigenmodes. We see that most all of the lower modes are well fit, while the higher modes contribute hundreds of units of "$\chi^2$" to the $\xi$ statistic. In sum,

- Minimizing $\xi$ can lead to pathological fits

- Even a "good" fit may have artificially large $\chi^2$

There are several cures to the problem of an unstable covariance matrix in small samples. Two unsatisfactory ones are:

- Increase the statistics several-fold

- Fit to points in a smaller time range

Two more promising choices are

- Fit the covariance matrix to a parametric form, e.g. the free field form for the appropriate four-point function.

- Fit to a smaller number of eigenmodes, which are reasonably well determined

We have opted for the latter method, truncating the observed covariance matrix to a small (typically 6-10) number of modes before inverting and defining a statistic

$$\xi_{(M)} \equiv \frac{N-M}{N-1} \sum_{\alpha=1}^{M} \frac{1}{\lambda_\alpha} (\vec{e}_\alpha \cdot \delta G)^2 \qquad (2)$$

where $N$ is the sample size and $M$ is the number of modes kept. For $M = T$ this of course is the usual F-statistic. One can verify this does something reasonable in a simple Gaussian model of our data. We construct a multi-normal distribution with zero mean and with covariance taken from typical real data. Drawing samples of N vectors from this sample, we then ask for the $\xi$ of the point $\vec{0}$, i.e. the true mean. In figure 3, we plot the confidence levels for such fits, i.e. the fraction of samples which have worse $\xi$ as a function of the reduced $\xi$. For the parameters shown in the figure, the conventional SVD goodness-of-fit statistic is large, and if interpreted as a $\chi^2$ would lead to the overwhelming rejection of the hypothesis that $\vec{0}$ is the true mean at most any confidence level. By contrast, the mode-truncated statistic appears to be close to a $\chi^2$. In practice we take advantage of this and *assume* it is a $\chi^2$ when assigning confidence levels our fits. Of course it would be better to generate the correct distribution, but the difference is not crucial.

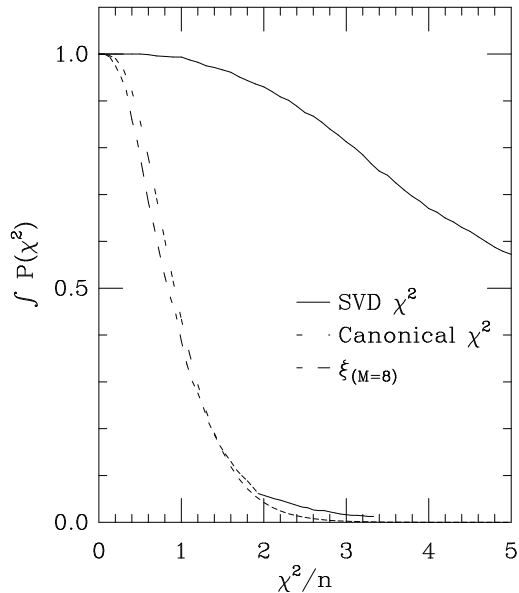

Figure 3. Comparison of the conventional SVD $\chi^2$ statisitic and the reduced mode statistic, using a realistic spectrum of eigenvalues.

As a practical matter, one must choose how many modes to keep by studying the effect on the bottom-line fit parameters. As shown in figure 4, the dependence is typically very weak until too many modes are added and the confidence level drops to near zero, a which point we reject the fit anyway. Based on such studies we typically choose 6-10 modes depending on the channel and fit range.

Secondly one must choose a fit model. We would have preferred to fit to the entire time domain, using several particles in each channel, or including some perturbative model of the short time behavior of our correlation functions. However, because we use extended (wall) sources we cannot claim to know the short-time behavior. Further, since we use Landau gauge, our wall sources evidently create correlations which can in part be understood at the independent propagation of quark states. To model this early time behavior properly would require too many parameters. By using simulated annealing and other minimization techniques, our fitting routines can handle models with a few (e.g. three) poles, but even this is not enough. Accordingly we chose to stick to single particle fits, and used the following protocol:

- For each time $t_{\min}$ perform fits the region $t_{\min}$-$t_{\text{last}}$ varying $N_{\text{mode}}$ between 6 and 12.

- Find the smallest $t_{\min}$ for which a "reasonable" (CL $> 0.2$) fit can be found.

- Throw away one timeslice and one mode for the definitive fit.

This fitting was performed automatically, and statistical errors were assessed using 250 bootstrap samples. In every case quoted, the errors were essentially symmetric, so we give only a single number for the error.

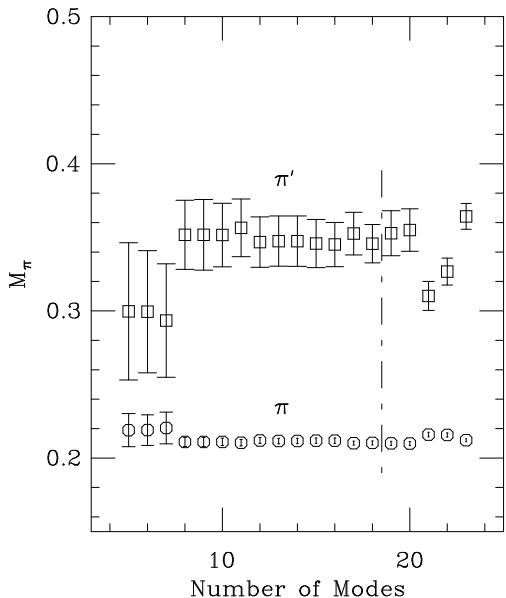

Figure 4. Mass parameters in a two particle fit to the $\pi$ correlator as a function of the number of modes kept. Fits to the right of the dashed line have unacceptable confidence levels.

## 3. Results

The main results are summarized in Table 2, where we show various parameters extrapolated to the chiral limit. The first two rows should be compared to the physical values of 1.22 and 10.1 respectively. Evidently the bottom line does not look too bad at $\beta = 6.4$. The parameter $\Delta$ is the mass of the the non-Goldstone pions in the chiral limit, obtained by fitting $m_\pi^2 = \Delta^2 + Am_q$. The ratio $\Delta/m_\rho^2$ remains fairly constant independent of $\beta$, as expected for a particle whose mass arises from staggered flavor symmetry breaking effects of order $\mathcal{O}(a^2)$.

Table 2
Results extrapolated to $m_q = 0$. Errors are derived from a 250 sample bootstrap

| $\beta$ | 6.0 | 6.2 | 6.4 |
|---|---|---|---|
| $N/\rho$ | $1.44 \pm .05$ | $1.37 \pm .03$ | $1.23 \pm .05$ |
| $N/f_\pi$ | $11.0 \pm .6$ | $11.3 \pm .4$ | $10.7 \pm .6$ |
| $\Delta/m_\rho^2$ | $.88 \pm .04$ | $.84 \pm .03$ | $.76 \pm .07$ |
| $a^{-1}(m_\rho)$ | $1.91 \pm .06$ | $2.61 \pm .05$ | $3.39 \pm .11$ |
| $a^{-1}(m_N)$ | $1.62 \pm .07$ | $2.32 \pm .06$ | $3.41 \pm .11$ |
| $a^{-1}(f_\pi)$ | $1.75 \pm .06$ | $2.64 \pm .07$ | $3.55 \pm .16$ |

## Acknowledgement
These calculations were performed on the Crays at the National Energy Research Supercomputer Center.